\documentclass[letterpaper]{article} 
\usepackage{aaai23}  
\usepackage{times}  
\usepackage{helvet}  
\usepackage{courier}  
\usepackage[hyphens]{url}  
\usepackage{graphicx} 
\usepackage{multirow}
\usepackage{natbib}  
\usepackage{caption} 
\usepackage{algorithm}
\usepackage{algorithmic}

\usepackage{newfloat}
\usepackage{listings}
\DeclareCaptionStyle{ruled}{labelfont=normalfont,labelsep=colon,strut=off} 
\floatstyle{ruled}
\newfloat{listing}{tb}{lst}{}
\floatname{listing}{Listing}
\title{Practical Lessons on Optimizing Sponsored Products in eCommerce}
\author{
    Yanbing Xue,\textsuperscript{\rm 1}
    Bo Liu,\textsuperscript{\rm 1}
    Weizhi Du,\textsuperscript{\rm 1}
    Jayanth Korlimarla,\textsuperscript{\rm 1}
    Musen Wen \textsuperscript{\rm 1}
}
\affiliations{
    \textsuperscript{\rm 1}Walmart eCommerce\\

    850 Cherry Ave\\
    San Bruno, California 94066 USA\\
    \{yanbing.xue, bo.liu, weizhi.du, jayanth.korlimarla, musen.wen\}@walmart.com
}

\usepackage{bibentry}

\begin{document}

\maketitle

\begin{abstract}
In this paper, we study multiple problems from sponsored product optimization in ad system, including position-based de-biasing, click-conversion multi-task learning, and calibration on predicted click-through-rate (pCTR). We propose a practical machine learning framework that provides the solutions to such problems without structural change to existing machine learning models, thus can be combined with most machine learning models including shallow models (e.g. gradient boosting decision trees, support vector machines). In this paper, we first propose data and feature engineering techniques to handle the aforementioned problems in ad system; after that, we evaluate the benefit of our practical framework on real-world data sets from our traffic logs from online shopping site. We show that our proposed practical framework with data and feature engineering can also handle the perennial problems in ad systems and bring increments to multiple evaluation metrics.
\end{abstract}

\section{Introduction}

In recent years, the world has witnessed remarkable increase in the number and quality of new CTR models built in eCommerce systems. The main factor contributing to this progress and improvement is the amount of computing power available to train more and more complex models. However, this improvement may not be possible when the computing power is under limited quota or when a fast-track improvement is preferred without structural change to the models. To alleviate this problem, we study various ways to improve the quality of CTR models by solving multiple ubiquitous perennial problems, while keeping using the same model without touching its structure. Our specific focus in this work is on the solution of position-based de-biasing, click-conversion multi-task learning, and calibration on predicted click-through-rate (pCTR).

One problem of CTR models in ad system is that the user-item interactions (e.g. views and clicks) are remarkably affected by the positions this ad item is displayed. Based on our beforehand observation from our online shopping site, as the position increases, the CTR drops significantly. Such observation indicates that more clicks on an ad item is not necessarily equal to the preference of this item from users, it may be just because this ad item is in a good position. Therefore, the the training data collected from past traffic logs contains position-based bias. One effective way to handle such position-based bias is to model it in a separate tower in deep learning models. For example,  \cite{10.1145/3298689.3347033} proposed a framework that models position-based bias in a separate tower as the factor that biases CTR. When training the model, the click label obtained from past traffic logs is considered biased as the bias factor times pCTR from CTR tower; when making unbiased predictions, the bias factor is set aside, and only pCTR from CTR tower is considered. Another framework \cite{10.1145/3298689.3346997} models the position-based bias as a term in addition to pCTR from CTR tower. In our paper, to leave the model structure unchanged, we model the position-based bias as weights of instances in training data. When making unbiased predictions, the position-based instance weight is set aside, and each instance is weighted equally for unbiased pCTR. Apart from instance weights, we also propose a coarse position-based de-biasing technique via feature engineering. In real-world websites and apps, ad items cannot be displayed in one granularity. For example, we need to scroll down or right into a new batch of more ad items. In such case, we propose a new granularity of history-based features on the batch level - that is - we calculate the historical views/clicks/CTR of each ad item on each batch. Such batch-level history-based features considers bias from positions since former batches typically have higher feature values than latter batches, and are also available for predictions.

Another problem of CTR models in ad system is its CTR-oriented nature. Traditional optimization techniques of CTR models do not consider conversions, thus possibly leading to an outcome that CTR is increased while CVR (click-conversion rate) and CTCVR (view-conversion rate), which may undermine the benefit of advertisers and in long run also ad platform since the advertisers will become unwillingly for price auctions. Therefore, one solution to handle such problem is multi-task training considering both clicks and conversions. One framework \cite{10.1145/3209978.3210104} was proposed for CVR models, where the CVR model is trained on CTR and CTCVR labels following the equation that $\texttt{CTCVR}=\texttt{CTR}\times\texttt{CVR}$. Another framework \cite{10.1145/3219819.3220007} was proposed to combine CTR and CVR training using multiple expert towers and multiple gating functions (one for each expert-task pair). In this paper, to leave the model structure unchanged, we propose two simple methods on training data to combine CTR and CVR training. One is stochastic label aggregation where we randomly keep one label between CTR and CTCVR label; the other is instance weight, where we reduce the weight of instances that are clicked but not converted.

The third problem of CTR model resides in output calibration. Traditional calibration techniques, including Platt's scaling and isotonic regression, are engaged in model training and cannot be changed real-time. To enable the real-time capability of calibration change, we proposed a simple but effective technique from price squashing, which adds an exponent to CTR value. Briefly, the larger exponent, the more CTR is emphasized in calibration.

In the next few sections, we elaborate our practical framework on multi-granularity de-biasing, multi-task CTR model, and price squashing for ad quality control, respectively. Then we conduct multiple experiments on our online shopping sites with real-world traffic logs showing the effectiveness our our practical framework.

\section{Multi-Granularity De-Biasing}
De-biasing is a hot topic of ad systems aimed to handle bias hidden in ads when predicting CTR. Such bias is typically introduced by ad item position - when an ad item is displayed in a bucket with a large view counts, users tend to only view the ads displayed in the front while ignoring those in the rear. In this section, we propose a practical multi-granularity de-biasing framework for modern online shopping website. Such framework owns the following advantages: (1) it implements position de-biasing in multiple granularities, which is specially curated for modern online shopping websites; (2) it leaves the model structure untouched and thus compatible with most CTR models.

\subsection{Refined non-linear de-biasing}\label{ref:pos_debias}
In traditional de-biasing frameworks, the attributes regarding de-biasing - say ad item position - are treated as common features when training a CTR model. The pCTR is predicted with these attributes pre-filled with a fixed value since the position is unknown.

There are multiple problems with the traditional de-biasing frameworks: (1) pre-filled values of the attributes lead to biased predictions; (2) bias-related attributes may affect pCTR in a different way than common features. To handle these problems, a more recent Position-bias Aware Learning (PAL) framework \cite{10.1145/3298689.3347033} proposed a two-tower model. Basically, this model splits the biased CTR obtained from historical logs into two factors:
$$P(y=1|\textbf{x},\texttt{pos})=P(v=1|\texttt{pos})P(y=1|\textbf{x},v=1)$$
by assuming: (1) the probability that an ad item has been viewed ($v=1$) is only dependent on the probability that the associated position has been observed; (2) the probability that an ad item is clicked is independent on its position if it has been viewed.

The advantage of PAL framework is that it decouples the ad item position and other features into two towers via conditional independence, yet one problem also arises: the CTR model needs changing structurally. Such changes needs to be curated model-to-model and is hardly feasible for shallow models (gradient boosting decision trees, support vector machines, etc).

To solve the problem of PAL framework, we propose a universal position-based de-biasing framework without touching model structure as follows:

\begin{figure}[t]\centering
\includegraphics[width=\columnwidth]{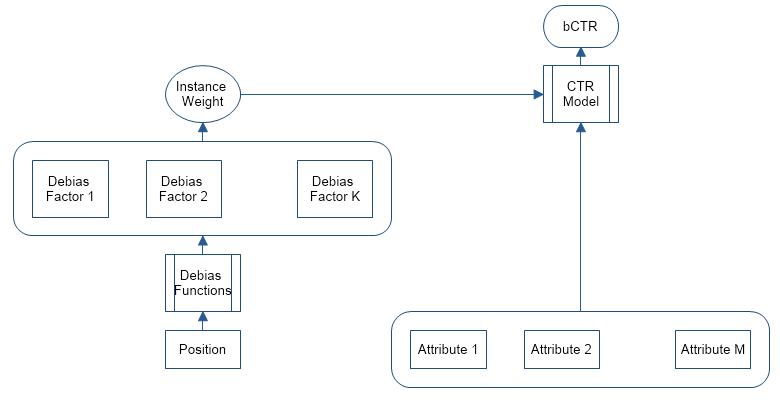}
\caption{Flowchart of Training a De-Biasing Model}
\label{fig:debias_train}
\end{figure}

(1) For training, we generate a set of de-bias functions regarding ad item position. Basically, these functions should be incremental regarding ad item position since labels from latter positions are less biased and should be emphasized. The common choices include polynomial, logarithmic, and inverse propensity weighting functions ;

(2) Then we generate instance weight from the linear combination of de-bias functions with ad item position. Such instance weight will be considered for CTR model training;

(3) When Predicting, we will not generate instance weight since we would like to obtain the unbiased $pCTR$ when the ad item position is unknown.

\begin{figure}[t]\centering
\includegraphics[width=.75\columnwidth]{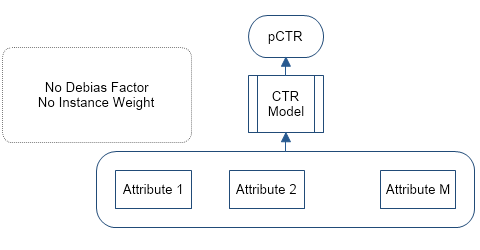}
\caption{Flowchart of Predicting a pCTR}
\label{fig:debias_test}
\end{figure}

In this way, we (1) achieve the position-based de-biasing with CTR model structure untouched, (2) maintain the non-linearity of de-biasing factors similar to PAL framework, and (3) circumvent the bias introduced by the pre-filled values of ad item position.
\subsection{Coarse bucket-level de-biasing}
In modern online shopping websites, ad items can be displayed in multiple granularities. A Typical example is \texttt{Walmart.com}: ad items among the search results are displayed in certain rows of a 2-d grid, while ad items below the search results are displayed in a rollable ribbon.

\begin{figure}[t]\centering
\includegraphics[width=\columnwidth]{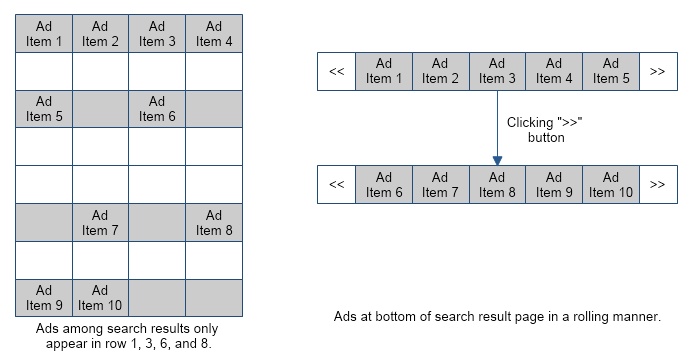}
\caption{Ad item display in search result page of \texttt{Walmart.com}}
\label{fig:seach_ingrid_bottom}
\end{figure}

According to our before-hand study, the coarse ad item position will also determine whether an ad item is likely viewed by a user. For ad items among the search results, those in the same row are similarly viewable to users, while ad items in latter rows are less viewable than those in former rows. For ad items below the search results, ad items in the same batch (e.g. ad item 1$\sim$5 in Fig \ref{fig:seach_ingrid_bottom}) are similarly viewable to users, while ad items in latter batches are less viewable than those in former batches. In other words, the position-based bias not only exists in each ad item position, but also in each coarse level of ad item positions, which we name as a ``bucket''.

Due to the bias existing in each bucket of ad item positions, we introduce a family of features: \texttt{Bucket-wise History-based Features}. Briefly, these features are historical metrics (clicks, views, CTR) of an ad item at bucket levels (e.g. rows of ads among search results, batches of ads below search results). More precisely, in addition to traditional history-based features at page level (e.g. search result page) and module level (e.g. among search results, below search results), we also extract bucket-level history-based features for de-biasing and performance boosting purpose.

The advantages of bucket-level history-based features reside in both performance and de-biasing: bucket-level history-based features (1) provide more refined information than traditional history-based features, thus improving model performance, (2) circumvent sparsity problem of more refined history-based (e.g. position-based) features, thus guaranteeing better availability, (3) provide bucket-level information for de-biasing purpose, (4) are available even when making predictions, thus circumventing the unavailability of ad item position from traditional de-biasing settings, and (5) are compatible with other position-based de-biasing frameworks (e.g. our non-linear de-biasing in Section \ref{ref:pos_debias}, PAL).
\section{Multi-Task CTR Model}
In traditional CTR models, the label of the training data is only focused on whether this ad item was clicked or not. Such setting can guarantee the experience of the users, since the probability of the user clicks is maximized. However, such setting may not guarantee the revenue of the advertisers, since users may just click these ad items without purchasing (converting) them, which is not beneficial to advertisers and may also undermine the well-being of the ad platform since the advertiser may become unwilling put high bids on the ad platform in the long run. To relieve such problem, a viable solution is to construct a multi-task CTR model which also takes conversion (CVR) into consideration. General multi-task learners, say Multi-gate Mixture-of-Experts (MMoE) \cite{10.1145/3219819.3220007}, can be an option. However, these general multi-task learner ignores the proximity between clicks and conversions: conversions are prerequisites of clicks. In other words, conversions indicate clicks, and non-clicks indicate non-conversions. Another disadvantage is the structural change to existing CTR models.

To relieve these problems, we propose a practical multi-task ranker combining CTR and CVR labels together. The combination of CTR and CVR labels can be achieved in two options:
\subsection{Stochastic label aggregation}
One option to combine CTR and CVR labels is stochastic label aggregation. Briefly, an aggregated label is generated using the following rules: (1) When an ad item is not clicked, it is not converted either, and the aggregated label will be 0; (2) when an ad item is both clicked and converted, the aggregated label will be 1; (3) when an ad item is clicked but not converted, there is a probability of $a\%$ the aggregated label will become 0 and a probability of $(1-a\%)$ the label will become 1, where $a\%$ is a tunable parameter. By applying such aggregated labels for training, ad items that are frequently clicked but rarely converted tend to have a lower pCTR than those frequently converted.
\subsection{Instance weighting}
Another option to combine CTR and CVR labels is instance weighting. Briefly, an instance in training data is weighted in the following manner: when an ad item is clicked but not converted, the weight of this ad item for training is dimmed into $a\%$ of the weight of converted ad items, where $a\%<1$ is a tunable parameter. By applying such changed weights for training, ad items that are frequently clicked but rarely converted tend to have a lower impact to the ranking model than those frequently converted, thus having a lower pCTR.

\section{Price Squashing in Ad Quality Control}
Calibration is a commonly-used strategy to handle the inaccuracy of pCTR in ad systems. Popular strategies for calibration are typically ML models include isotonic regression, Platt's scaling, etc. However, the main disadvantage of these strategies is the nature of ML models - change of calibration strategies typically requires retraining of ML models and cannot be finished real-time.

\begin{table*}[t]\centering
\begin{tabular}{|c|c|c|c|c|}
\hline
Method$\downarrow$ & CTR AUROC & CTR AUPRC & CTCVR AUROC & CTCVR AUPRC\\
\hline
Raw & 0.7522 & 0.7493 & 0.7054 & 1.43$\times 10^{-3}$\\
\hline
\textbf{Prac} & \textbf{0.7585} & \textbf{0.7508} & \textbf{0.7221} & \textbf{1.51$\times$10$^{-3}$}\\
\hline
AbDeb & 0.7582 & 0.7502 & 0.7125 & 1.45$\times 10^{-3}$\\
\hline
AbMul & 0.7545 & 0.7495 & 0.7189 & 1.48$\times 10^{-3}$\\
\hline
\end{tabular}
\caption{Offline performance of all methods for ads among search results.}
\label{Table:middle}
\end{table*}

\begin{table*}[t]\centering
\begin{tabular}{|c|c|c|c|c|c|c|c|c|}
\hline
Method$\downarrow$ & CTR AUROC & CTR AUPRC & CTCVR AUROC & CTCVR AUPRC\\
\hline
Raw & 0.7670 & 0.7274 & 0.7064 & 1.13$\times 10^{-3}$\\
\hline
\textbf{Prac} & \textbf{0.7834} & \textbf{0.7349} & \textbf{0.7459} & \textbf{1.34$\times$10$^{-3}$}\\
\hline
AbDeb & 0.7825 & 0.7347 & 0.7225 & 1.25$\times 10^{-3}$\\
\hline
AbMul & 0.7805 & 0.7341 & 0.7312 & 1.27$\times 10^{-3}$\\
\hline
\end{tabular}
\caption{Offline performance of all methods for ads below search results.}
\label{Table:bottom}
\end{table*}

To enable the real-time capability of calibration, we borrow the idea of price squashing. Price squashing is general strategy in ads marketplace auction. In ad system, the final ranking of ad items is generated regarding pCTR $\times$ CPC, where CPC is cost per click. By introducing price squashing, the final ranking criteria now becomes pCTR$^c\times$ CPC, where $c$ is a tunable parameter. When $c>1$, the final ranking criteria focuses pCTR over CPC, and vice versa. Clearly by changing $c$, the new final ranking criteria will take effect real-time.
\section{Experiments and Results}
In this section, we test our approach on real-world data both offline and online. The real-world data set extracted directly from full traffic of ad service logs from \texttt{Walmart.com}.

\subsection{Offline experiment}
\subsubsection{Experimental settings}
The data sets in offline experiments consists of full traffic of ad service logs from \texttt{Walmart.com} regarding ads among search results and below search results. In each placement, the feature vector of each instance contains up to 30 features regarding the pair of search query and ad item viewed (NOT only impressed), the labels of each instance contain CTR (view-click) label and CTCVR (view-conversion) label. Both data sets contain full traffic from the identical consecutive 30-day window. The training sets only contain full traffic from the former 23 days, while the testing sets only contain full traffic from the latter 7 days.

The following methods are being used to demonstrate the benefits of our practical framework for ad system:

\textit{Raw}: A raw GBDT CTR model without de-biasing, multi-task, or price squashing capabilities;

\textit{Prac}: Our practical framework with multi-granularity de-biasing, click-conversion multi-task learning, and price squashing capabilities;

\textit{AbDeb}: For ablation test purpose. Our practical framework with multi-granularity de-biasing and price squashing capabilities;

\textit{AbMul}: For ablation test purpose. Our practical framework with click-conversion multi-task learning and price squashing capabilities.

We evaluate the performance of the different methods by calculating the area under ROC curve and precision-recall curve (AUROC and AUPRC) all the methods achieve on the test data. The learning considers the training data only, and the two aforementioned evaluation metrics are always calculated on the test set.

\subsubsection{Experimental results}

Table \ref{Table:middle} and Table \ref{Table:bottom}\footnote{Due to security concerns, we add an undisclosed random number to each evaluation metric.} show the benefit of our practical framework with multi-granularity de-biasing, click-conversion multi-task learning, and price squashing capabilities both individually and jointly:

\noindent\emph{Effect of multi-granularity de-biasing}: On both data sets, {\em AbDeb} outperforms {\em Raw}; {\em Prac} outperforms {\em abMul}. These two groups of comparisons show multi-granularity de-biasing improves the learning performance when compared with models without de-biasing at the same training data.

\noindent\emph{Effect of click-conversion multi-task learning}: Also, on both data sets, {\em AbMul} outperforms {\em Raw}; {\em Prac} outperforms {\em AbDeb}. These two groups of comparisons show click-conversion multi-task learning improves the performance. Again the results are matched at the same training data.

\noindent\emph{Effect of our practical framework}: Overall, on both data sets, our framework {\em Prac}, which is the combination of multi-granularity de-biasing, click-conversion multi-task learning, and price squashing, achieved the highest performance. This supports and confirms the effectiveness of our practical framework with multi-granularity de-biasing, click-conversion multi-task learning, and price squashing capabilities;

\begin{table}[t]\centering
\begin{tabular}{|c|c|c|}
\hline
Method$\rightarrow$ & \textbf{Prac} & Raw\\
\hline
eCTR & \textbf{1.023} & 1\\
\hline
eCPMV & \textbf{1.033} & 1\\
\hline
ROAS & \textbf{1.074} & 1\\
\hline
Distinct Sellers & \textbf{1.017} & 1\\
\hline
Distinct Ad Items & \textbf{1.161} & 1\\
\hline
\end{tabular}
\caption{Online performance of our practical framework.}
\label{Table:online}
\end{table}

\subsection{Online experiment}
\subsubsection{Experimental settings}
Apart from offline experiment, we also conducted an online experiment, where we put our practical framework with multi-granularity de-biasing, click-conversion multi-task learning, and price squashing capabilities online for ads among search results at \texttt{Walmart.com}. We started from 1\% of full traffic, and gradually increased the traffic to 30\% and kept for 14 days. Then we compare multiple evaluation metrics of our practical framework (\textit{Prac}) with current production model (\textit{Raw}).

\subsubsection{Experimental results}
In online experiment result in \ref{Table:online}\footnote{Due to security concerns, we can only provide fractions of all metrics between our practical framework and production model.}, our framework {\em Prac}, which is the combination of multi-granularity de-biasing, click-conversion multi-task learning, and price squashing, achieved higher performance on all evaluation metrics, including expected click-through rate (eCTR), expected cost per thousand views (eCPMV), return on ad spend (ROAS), distinct sellers displayed, and distinct ad items displayed.

\section{Conclusion}
Position-based de-biasing, click-conversion (CTR-CVR) multi-task learning, and predicted click-through-rate (pCTR) calibration are perennial problems in ad systems. In this work, we proposed a practical learning framework for learning a CTR model that handles these problem via data and feature engineering without structural change to existing models. The main advantages of our learning framework is compatibility with most machine learning models and zero increment on training time consumption. Our experimental results also show that our learning framework is able to handle the aforementioned problems and bring increments to multiple metrics in ad system.
\bibliography{sample-base}
\end{document}